\documentclass[a4paper,12pt]{article}
\date{}
\begin{document}
\renewcommand{\thefootnote}{\fnsymbol{footnote}}
\begin{center}
{\Large\bf Meson mass spectrum and OPE: matching to the large-$N_c$
QCD}\footnote{Talk at V-th International Conference
``Quark Confinement and the Hadron Spectrum'', Gargnano, Garda Lake, Italy,
10-14 September 2002.}\footnote{\uppercase{T}his work is supported by \uppercase{G}rant
\uppercase{RFBR} 01-02-17152, \uppercase{INTAS} \uppercase{C}all 2000 
grant
(\uppercase{P}roject 587), \uppercase{R}ussian \uppercase{M}inistry
of \uppercase{E}ducation grant \uppercase{E}00-33-208, the
\uppercase{P}rogram "\uppercase{U}niversities of
\uppercase{R}ussia: \uppercase{F}undamental \uppercase{I}nvestigations"
(grant \uppercase{UR}.02.01.001). S.A. is also 
supported by \uppercase{G}rant \uppercase{INTAS-YS} 4384 and 
\uppercase{G}rant for
young scientists of \uppercase{S}t.\uppercase{P}etersburg
\uppercase{M}02-2.4\uppercase{K}-15.
}
\vspace{1cm}

{\large\bf A.~A. Andrianov}\\
St.Petersburg State University and INFN, Sezione di Bologna\\
\medskip

{\large\bf V.~A. Andrianov} and {\large\bf S.~S. Afonin}\\
St.Petersburg State University
\end{center}


\abstract{The relations
between masses and decay constants of variety of meson resonances
in the energy range 0--3 GeV are verified from the string-like, 
linear mass spectrum for vector,
axial-vector, scalar and pseudoscalar mesons with a  universal slope. 
The way to match the universality with the Operator
Product Expansion (OPE) is proposed. The necessity of small deviations from 
linearity in parameterization of the meson mass spectrum and their 
decay constants is proven from matching to OPE.}

\vspace{1cm}

As it follows from phenomenology \cite{bgs,ani2}  the
masses squared of mesons of a given spin 
increase linearly  with the number of (radial) excitation $n$ 
which can be thought of as a hint on the string structure of QCD.
The string picture gives an equal slope of these 
trajectories for different quarkonium mesons
since this quantity is proportional
to the string tension depending on gluodynamics only.
In our talk  we examine possible corrections to the
linear trajectories for vector~(V), axial-vector~(A), scalar~(S),
and pseudoscalar~(P) case. 
Our method is based on the consideration of two-point correlators
of quark currents in the large-$N_c$ limit of QCD \cite{hoof}. On one
hand, by virtue of confinement they are saturated by an infinite set of
narrow meson
resonances, on the other hand their high-energy asymptotics are
provided by the perturbation theory and the OPE \cite{svz} with condensates.

We adopt the following ansatz for the meson mass spectrum 
(compare to \cite{beane}):
\begin{equation}
\label{acor 1}
m_R^2(n)=m_{0,R}^2+a\,n+\frac{d_R}{n+1}\,,
\end{equation}
where
$R\equiv V,A,S,P$ and the corrections to linear trajectory fit a
possible deviation from the string picture in QCD. It turns out
that for the consistency with the OPE one needs the following
conditions on the residues:
\begin{equation}
\begin{array}{rcl}
Z_{V\!\!A}(n)&=&t\left(m_{V\!\!A}^2(n)\right)\frac{dm_{V\!\!A}^2(n)}{dn}\,,
\\[4pt]
Z_{S\!P}(n)&=&t\left(m_{SP}^2(n)\right)
\left(m_{S\!P}^2(n)-\frac{3\alpha_s}{4\pi^3C}\lambda^2\right)
\frac{dm_{S\!P}^2(n)}{dn}\,;
\end{array}\label{conVA}
\end{equation}
\begin{equation}
\label{t1}
t\left(m_R^2(n)\right)=C+\sum_{i>0}A_i\exp\left(-B_i
m_R^2(n)\right)\,, \qquad B_i>0\,,
\end{equation}
where $\lambda$ represents
the possible dimension two condensate. The wave 
function corrections  in Eq.~(\ref{t1}) are taken in the exponential form 
in order to avoid positive 
powers of logarithms of momentum in the high-energy 
asymptotics in accordance with OPE.

Our numerical calculations (see the tables at the end) have shown
that:\\
1. The linear ansatz ($d_R=0$) for the
meson mass spectrum with a universal slope 
fits rather well the meson phenomenology \cite{ani2} but
 for the corresponding residues (wave
function moments) a related constant or linear ansatz ($A_i=0$)
in Eq.~(\ref{conVA}) leads to a striking 
disagreement with the OPE and, in particular, does not provide 
Chiral Symmetry Restoration (CSR) \cite{av} , i.e. fast decreasing of differences 
between correlators of parity doublers ($S - P,\,V- A$)
(compare to \cite{p1}).\\
2. Nonlinear
corrections ($d_R\neq 0$) 
to the meson trajectories with universal slope may bring CSR but they do not
improve drastically the matching to the OPE in each channel.\\
3. The OPE
matching may be fulfilled with the help of exponentially
decreasing corrections ($A_i\neq 0$) to the
meson residues (wave function momenta), Eq.~(\ref{t1}) 
(in our calculations we introduced 
two of such terms in each channel).
They control
substantially the values of condensates in high-energy asymptotics though 
being numerically negligible at low
and intermediate energies. 

We express our gratitude to the organizers of the
"5-th International Conference Quark Confinement and the Hadron Spectrum"
for hospitality.


\begin{table}
\caption{An example of meson mass spectra. 
The inputs are: $a=(1200\,\mbox{MeV})^2$,
$\langle\bar qq\rangle=-(240\,\mbox{MeV})^3$,
$\langle\left(G_{\mu\nu}^a\right)^2\rangle=(360\,\mbox{MeV})^4$,
$\lambda=0$ (this condensate has a tiny influence on the results),
$f_{\pi}=87\,\mbox{MeV}$, $\alpha_s=0.5$. The units are:
$m(n)$, $F(n)$, $m_0$ --- MeV; $d$ --- MeV${}^2$; $A_i$ --- MeV${}^0$;
$B_i$ --- MeV${}^{-2}$. The corresponding experimental values \cite{pdg}
are displayed in brackets.\vspace*{1pt}}

\bigskip

{\footnotesize
\begin{tabular}{|c|c|c|}
\hline
{} &{} &{} \\[-1.5ex]
 Case & Inputs & Predictions\\[1ex]
\hline
{} &{} &{} \\[-1.5ex]
& $m(0)=770\,(769.3\pm0.8)$ & $m_0=900$, $d=-470^2$ \\
   $V$ & $m(1)=1460\,(1465\pm25)$ & $m(2)=1900\,(?)$,
   $m(3)=2250\,(2149\pm17)$, \dots \\
   & $F(0)=150\,(154\pm8)$ & $A_1=7\cdot10^{-4}$,
   $A_2=-8\cdot10^{-10}$ \\
   && $B_1=1000^{-2}$, $B_2=3160^{-2}$ \\
  \hline
   & $m(0)=1200\,(1230\pm40)$ & $m_0=910$, $d=790^2$ \\
   $A$ & $m(1)=1600\,(1640\pm40)$ & $m(2)=1970\,(?)$,
   $m(3)=2290\,(?)$, \dots \\
   & $F(0)=130\,(123\pm25)$ & $A_1=-2$,
   $A_2=10^{-8}$ \\
   && $B_1=1730^{-2}$, $B_2=14140^{-2}$ \\
  \hline
   & $m(0)=900\,(980\pm10)$ & $m_0=910$, $d=-140^2$ \\
   $S$ & $m(1)=1500\,(1500\pm10)$ & $m(2)=1920\,(?)$,
   $m(3)=2260\,(2197\pm17)$, \dots \\
   & $F(0)=320$ & $A_1=22$,
   $A_2=-1$ \\
   && $B_1=2650^{-2}$, $B_2=1730^{-2}$ \\
  \hline
   & $m(0)=0\,(\approx135)$ & $m_0=720$, $d=-720^2$ \\
   $P$ & $m(1)=1300\,(1300\pm100)$ & $m(2)=1790\,(1801\pm13)$,
   $m(3)=2160\,(?)$, \dots \\
   $\pi$ - on & $F(0)=400$ & $A_1=0.31$,
   $A_2=-0.27$ \\
   && $B_1=1730^{-2}$, $B_2=2240^{-2}$ \\
  \hline
   & $m(0)=1300\,(1300\pm100)$ & $m_0=1390$, $d=-490^2$ \\
   $P$ & $m(1)=1800\,(1801\pm13)$ & $m(2)=2170\,(?)$,
   $m(3)=2480\,(?)$, \dots \\
   $\pi$ - out & $F_\pi=400$ & $A_1=-26$,
   $A_2=690$ \\
   && $B_1=1730^{-2}$, $B_2=2240^{-2}$ \\
  \hline
\end{tabular}}
\medskip

\caption{The values of residues for the first four states.  In
brackets we show the corresponding value without exponential
corrections. Precision is $\pm5$~MeV.\vspace*{1pt}}
\medskip
\begin{center}
\begin{tabular}{|c|cccc|}
  \hline
  {} &{} &{} &{} &{}\\[-1.5ex]
  Case & $F(0)$ & $F(1)$ & $F(2)$ & $F(3)$ \\[1ex]
  {} &{} &{} &{} &{}\\[-1.5ex]
  \hline
  $V$ & $150\,(160)$ & $150\,(150)$ & $150\,(150)$ & $150\,(150)$ \\
  $A$ & $130\,(110)$ & $140\,(140)$ & $140\,(140)$ & $140\,(140)$ \\
  $S$ & $320\,(430)$ & $550\,(560)$ & $620\,(630)$ & $690\,(690)$ \\
  $P$ $\pi$-on & $400\,(-)$ & $540\,(530)$ & $620\,(620)$ & $670\,(670)$ \\
  $P$ $\pi$-out & $470\,(520)$ & $610\,(610)$ & $670\,(670)$ &
  $720\,(720)$ \\
  \hline
\end{tabular}
\end{center}
\end{table}

\end{document}